\documentclass[conference,letterpaper]{IEEEtran}

\usepackage{cite}

\ifCLASSINFOpdf
\else
   \usepackage[dvips]{graphicx}
   \graphicspath{{../figures/}}
   \DeclareGraphicsExtensions{.eps}
\fi
\usepackage[cmex10]{amsmath}
\usepackage{amssymb,amsthm}
\usepackage{dsfont}

\theoremstyle{remark}
\newtheorem{remark}{Remark}
\newtheorem{lemma}{Lemma}
\newtheorem{theorem}{Theorem}

\newtheorem{definition}{Definition}

\newtheorem{example}{Example}
\newtheorem{proposition}{Proposition}

\usepackage{psfrag}
\usepackage{array}
\usepackage{mdwmath}
\usepackage{mdwtab}
\usepackage{cases}
\usepackage{enumerate}

\usepackage{eqparbox}
\usepackage{mathtools}
\let\underbrace\LaTeXunderbrace

\usepackage{fixltx2e}
\usepackage{stfloats}

\newcommand{\xs}{x_{\text{s}} }
\newcommand{\xr}{x_{\text{r}} }
\newcommand{\xd}{x_{\text{d}} }

\newcommand{\widesim}[2][1.5]{%
  \mathrel{\overset{#2}{\scalebox{#1}[1]{\raisebox{-0.08cm}{$\sim$}}} }
}

\begin{document}
\title{Cooperative Relaying in a Poisson Field of Interferers: A Diversity Order Analysis}

\author{\IEEEauthorblockN{Ralph Tanbourgi, Holger J\"{a}kel and Friedrich K. Jondral}
\IEEEauthorblockA{Communications Engineering Lab, Karlsruhe Institute of Technology, Germany\\
Email: \{ralph.tanbourgi, holger.jaekel, friedrich.jondral\}@kit.edu}
}

\maketitle

\begin{abstract}
%\boldmath
This work analyzes the gains of cooperative relaying in interference-limited networks, in which outages can be due to interference and fading. A stochastic model based on point process theory is used to capture the spatial randomness present in contemporary wireless networks. Using a modification of the diversity order metric, the reliability gain of selection decode-and-forward is studied for several cases. The main results are as follows: the achievable \emph{spatial-contention} diversity order (SC-DO) is equal to one irrespective of the type of channel which is due to the ineffectiveness of the relay in the MAC-phase (transmit diversity). In the BC-phase (receive diversity), the SC-DO depends on the amount of fading and spatial interference correlation. In the absence of fading, there is a hard transition between SC-DO of either one or two, depending on the system parameters.\vspace{-.00cm}
\end{abstract}
\renewcommand\qedsymbol{\IEEEQEDopen}

\IEEEpeerreviewmaketitle
\vspace{-.1cm}
\section{Introduction}\label{sec:intro}
In spite of steadily increasing data rate demands, cooperative diversity---and most saliently, cooperative relaying---has emerged to a widely-recognized concept to increase reliability and/or throughput through exploration of spatial diversity. Cooperative relaying has gained practical relevance at least since its adoption in the 3GPP Rel-10 for 4G networks \cite{hoymann12}. Taking 4G as an example, the trend for networks goes toward interference-limitedness as they must cope with heterogeneity/coexistence, densification of devices and sometimes unpredictable deployments \cite{andrews13_1}. A better understanding of cooperative relaying in the presence of random interference is hence mandatory. Among the vast body of literature concerning relaying, most prominently \cite{cover79,laneman04}, there exist only a limited number of works that take into account the effect of random interference, see e.g., \cite{ganti12,altieri12}.

In the high reliability regime, the diversity order \cite{laneman04} metric can be used to measure the increase in robustness against random fluctuations in the channel. In the interference-free scenario, this regime is obtained by letting $\text{SNR}\to\infty$. Practically, this involves scaling the transmit power since the receiver noise cannot be lowered to an arbitrary extent. This observation, however, does not apply to interference-limited multi-user networks since jointly increasing transmit power does not increase the individual SIRs. This gives rise to the question of how to measure the diversity order of cooperative relaying in interference-limited networks appropriately. We propose a modified diversity order metric, namely \emph{spatial-contention} diversity order, which is based on scaling the density of active nodes in the network. We argue that this modified metric is more suitable for interference-limited networks since the spatial resource---which is considered the critical resource---is taken into account. Also, controlling the density of active transmissions has been understood as an important and effective means to increase network capacity, and is therefore the underlying mechanism of practical MAC protocols such as Aloha (spatial reuse with medium access probability) and CSMA (spatial inhibition of active nodes). It is hence worth studying the diversity behavior of cooperative relaying as a function of the density of active nodes. % Clearly, for studying the diversity order of cooperative techniques from a network's point of view, relating the density of active nodes to the high reliability regime may be the most appropriate way.

%the density of nodes plays an important role and may sometimes be the most accessible quantity to be controlled in highly dynamic environments.

Using point process theory, we derive a stochastic model to study the diversity of cooperative relaying in the presence of random interference. We aim at answering the following questions: How much diversity can we expect in the interference-limited case? How does spatial interference correlation and fading affect the achievable diversity gains?
\section{System Model and Assumptions}\label{sec:system_model}
To address the key questions of this work, we break the analysis down into a single snapshot of the network, in which a given transmission is interfered by randomly located nodes transmitting in the same time-frequency resource.

\subsection{Channel model}
%Among the different aspects of the system model, the channel model will be of importance as they will strongly influence the outcome of the analysis.
%\subsubsection{Path loss function}
The power path loss between two locations $x,y\in\mathbb{R}^2$ is given by the non-singular path loss function $\ell(|x-y|):=(1+|x-y|^{\alpha})^{-1}$, where $\alpha>2$ denotes the path loss exponent. % The use of a non-singular path loss function is crucial for a meaningful analysis, since the spatial correlation of the interference can be properly captured only by a non-singular path loss function \cite{ganti09}.
%\subsubsection{Channel fading}
Both the correlation and the statistics of the SIR strongly depend on the type of channel fading, and particularly on its distribution. Since the family of practical fading distributions is large, we focus on two extreme cases: frequency-flat block Rayleigh fading and path loss only, the former being usually considered as severe fading while the latter can be seen as the limiting case of weak scatterings.%\footnote{Clearly, other fading distributions, e.g., Rice fading, is in between the two extremes.}% As for the Rayleigh fading case, we will assume frequency-flat block fading.

\subsection{Relay protocol}
We consider a three-node configuration which consists of a source located at $x_{\text{s}}$, a destination located at $x_{\text{d}}$ and a half-duplex relay located at $x_{\text{r}}$. The locations $x_{\text{s}}$, $x_{\text{d}}$ and $x_{\text{r}}$ are arbitrary but fixed. Hence, we place the destination into the origin ($x_{\text{d}}=o$). %The relay operates in half-duplex mode due to practical constraints. 
The block is divided into two consecutive time slots over which the transmission takes place.

Selection decode-and-forward (SDF) \cite{laneman04} is used as the relay protocol. % for conveying information from source to destination. 
In SDF, the source broadcasts a packet in the first time slot, while the destination buffers what it receives and the relay tries to decode the packet. Depending on whether the relay was able to correctly decode the packet, either the relay or the source then re-transmits the packet to the destination in the second time slot. Finally, the destination appropriately combines the two copies prior to decoding the packet. %The fact that the destination re-transmits the packet whenever the relay fails to decode it renders this protocol adaptive and recovers the degree of freedom (second time slot) that would be lost otherwise.

\subsection{Interference model}
As the three-node configuration is part of a multi-user environment, it will suffer from interference from other transmitters (interferers). We assume that these interferers are distributed according to a stationary Poisson point process (PPP) with density $\lambda$.\footnote{Since the PPP assumption excludes any form of correlation in the nodes' locations, the considered source-relay-destination link is not \emph{typical}.} The PPP assumption is well-accepted for capturing the spatial randomness in contemporary wireless networks of several types \cite{weber10,andrews13_1}. Formally, we define
\begin{IEEEeqnarray}{rCl}
	\Phi:=\big\{(\mathsf{x}_{i},\mathsf{g}_{i},\mathsf{h}_{i}):\mathsf{x}_{i}\in\mathbb{R}^2,\,\mathsf{g}_{i}\in\mathbb{R}_{+},\,\mathsf{h}_{i}\in\mathbb{R}_{+}\big\},
\end{IEEEeqnarray}
where $\mathsf{x}_{i}$ denotes the random location of the $i$-th interferer, while the marks $\mathsf{g}_{i}$ and $\mathsf{h}_{i}$ define the channel fading gain from the $i$-th interferer to the relay and the destination, respectively. All marks are mutually i.i.d. and do not depend on the interferer locations. The intensity measure of $\Phi$ is given by
\begin{IEEEeqnarray}{rCl}
	&&\Lambda(A\times \Gamma\times\Upsilon):= \lambda\int_{A} \int_{\Gamma}\mathrm d\mathbb{P}\left(\mathsf{g}\leq g\right)\int_{\Upsilon}\mathrm d\mathbb{P}\left(\mathsf{h}\leq h\right)\,\mathrm dx\IEEEnonumber\\
	&&\;=\lambda|A|\mathbb{P}(\mathsf{g}\in\Gamma)\mathbb{P}(\mathsf{h}\in\Upsilon),\quad A\subseteq\mathbb{R}^2,\Gamma\subseteq\mathbb{R}_{+},\Upsilon\subseteq\mathbb{R}_{+}.\label{eq:intensity_measure}\IEEEeqnarraynumspace
\end{IEEEeqnarray}

\begin{remark}
For the path loss only model ($\mathsf{g}_{i}\equiv\mathsf{h}_{i}\equiv1\,\forall i$), the intensity measure reduces to $\lambda|A|$.  
\end{remark}
Thus, the interference at the relay and at the destination is\footnote{We use the short-hand notation $i\in\Phi$ instead of $(\mathsf{x}_{i},\mathsf{g}_{i},\mathsf{h}_{i})\in\Phi$.  Interference is treated as white noise. Without loss of generality, we set transmit power to one. We assume the interference power realizations to remain constant over the two considered time slots.}
%ame index convention as in \cite[Remark 2.2]{} for writing functionals of $\Phi$: we will sum over the indices $i$, i.e., $\sum_{i\in\Phi}$, rather than over the actual points $(\mathsf{x}_{i},\mathsf{g}_{i},\mathsf{h}_{i})$, i.e., $\sum_{(\mathsf{x}_{i},\mathsf{g}_{i},\mathsf{h}_{i})\in\Phi}$, for notational convenience.}
\begin{IEEEeqnarray}{rCl}
	\mathsf{I}_{\text{r}}:=\sum\limits_{i\in\Phi}\mathsf{g}_{i}\ell(|\mathsf{x}_{i}-\xr|)\quad \text{and} \quad
	\mathsf{I}_{\text{d}}:=\sum\limits_{i\in\Phi}\mathsf{h}_{i}\ell(|\mathsf{x}_{i}|).\IEEEeqnarraynumspace\label{eq:sn_def}
\end{IEEEeqnarray}
Note that $\mathsf{I}_{\text{r}}$ and $\mathsf{I}_{\text{d}}$ are correlated because of the common source of randomness given by the interferer locations $\{\mathsf{x}_{i}\}_{i=0}^{\infty}$.

\subsection{Performance metrics}
With the above setting, the SIR at the relay is given by
\begin{IEEEeqnarray}{rCl}
	\mathsf{SIR}_{\text{sr}}&&=\frac{\mathsf{u}_{\text{sr}}\ell(|\xs-\xr|)}{\mathsf{I}_{\text{r}}},
\end{IEEEeqnarray}
where $\mathsf{u}_{\text{sr}}$ denotes the channel fading gain on the source-relay link. %and $\eta$ is the inverse transmit SNR\@.
Given that the relay was able to decode successfully, the SIR at the destination after optimum combining is
\begin{IEEEeqnarray}{rCl}
	\mathsf{SIR}_{\text{srd}}&&=\frac{\mathsf{u}_{\text{sd}}\ell(|\xs|)+\mathsf{u}_{\text{rd}}\ell(|\xr|)}{\mathsf{I}_{\text{d}}},
\end{IEEEeqnarray}
where $\mathsf{u}_{\text{sd}}$ and $\mathsf{u}_{\text{rd}}$ are the channel fading gains on the source-destination and relay-destination links, respectively. When the relay fails to decode correctly, the transmitter re-transmits the packet and the SIR at the destination becomes
\begin{IEEEeqnarray}{rCl}
	\mathsf{SIR}_{\text{sd}}&&=\frac{2\mathsf{u}_{\text{sd}}\ell(|\xs|)}{\mathsf{I}_{\text{d}}}.
\end{IEEEeqnarray}
\begin{remark}
	When referring to the entire group of fading variables, i.e., $\{\mathsf{u}_{\text{sr}},\mathsf{u}_{\text{sd}},\mathsf{u}_{\text{rd}}\}$, $\{\mathsf{g}_{i}\}_{i=0}^{\infty}$ and $\{\mathsf{h}_{i}\}_{i=0}^{\infty}$, we will use the short-hand notation $\mathsf{u}$, $\mathsf{g}$ and $\mathsf{h}$, respectively. 
\end{remark}

In many cases the random fluctuations of the SIR cannot be tracked by the transmitter due to practical constraints, and particularly because the interference from many nodes cannot be known \emph{a priori}. %, since setting up and maintaining coordination links is often not feasible. 
This may lead to an outage, for which the probability of occurrence is a useful performance metric.
\begin{definition}
	The outage probability (OP) is defined as
	\begin{IEEEeqnarray}{rCl}
		q:=\mathbb{P}(\mathsf{SIR}<\beta)
	\end{IEEEeqnarray}
	for a pre-defined coding/modulation-specific threshold $\beta$.
\end{definition}
%In order to measure the reliability gain of diversity techniques the diversity order metric can be used. Traditionally, it involves letting the SNR tend to infinity, thereby marking the high reliability regime in the single-user case. Usually, letting the SNR tend to infinity implies increasing transmit power since receiver noise cannot be lowered to an arbitrarily extend \cite{}. But in a multi-user environment, there is no gain by having all nodes increasing their transmit power. So how do we get to the high reliability regime in the multi-user scenario?
We propose an alternative formulation of the diversity order metric that applies to a multi-user environment and which is based on controlling the density of simultaneous transmissions.
\begin{definition}
	The spatial-contention diversity order (SC-DO) is defined as
	\begin{IEEEeqnarray}{c}
		\Delta:=\lim_{\lambda\to 0} \frac{\log q}{\log \lambda}.
	\end{IEEEeqnarray}
\end{definition}

\begin{example}
	In the absence of the relay, the OP for Rayleigh fading is known to be \cite{baccelli06}
	\begin{IEEEeqnarray}{rCl}
		1-\exp\big\{-\lambda\pi^2\tfrac{2}{\alpha}|\xs|^{2}\beta^{\frac{2}{\alpha}}\csc(\tfrac{2}{\alpha}\pi)\big\}.
	\end{IEEEeqnarray}
	The SC-DO in this case is given by $\Delta=1$ as expected.
\end{example}

\section{Outage Analysis --- Rayleigh Fading}\label{sec:out_ray}
In most works, cooperative relaying is examined for the case of exponentially distributed fading gains with channel state information (CSI) available only at the receivers. We start our analysis by considering this scenario.

From \cite{laneman04}, the OP for SDF can be expressed as
\begin{IEEEeqnarray}{c}
	q=\underbrace{\mathbb{P}\hspace{-.03cm}\left(\mathsf{SIR}_{\text{sd}}\hspace{-.030cm}<\hspace{-.030cm}\beta,\,\mathsf{SIR}_{\text{sr}}\hspace{-.030cm}<\hspace{-.030cm}\beta\right)}_{:=q_{\text{BC}}}+\underbrace{\mathbb{P}\hspace{-.03cm}\left(\mathsf{SIR}_{\text{srd}}\hspace{-.030cm}<\hspace{-.030cm}\beta,\,\mathsf{SIR}_{\text{sr}}\hspace{-.030cm}\geq\hspace{-.030cm}\beta\right)}_{:=q_{\text{MAC}}},\IEEEeqnarraynumspace\label{eq:op_adaf}
\end{IEEEeqnarray}
where $q_{\text{BC}}$ and $q_{\text{MAC}}$ denote the OP in the Broadcast phase (BC-phase) and the MAC phase (MAC-phase), respectively. Treating these two expressions separately will be advantageous in the subsequent analysis. Applying stochastic geometry tools, \eqref{eq:op_adaf} can be calculated in semi-closed form.
\begin{proposition}\label{prop:rayray_op}
	Define
	\begin{IEEEeqnarray}{rCl}
	  &&\ell^{\ast}_{\text{sd}}(r):=\frac{1+|\xs|^{\alpha}}{1+r^{\alpha}},\quad\ell^{\ast}_{\text{rd}}(r):=\frac{1+|\xr|^{\alpha}}{1+r^{\alpha}},\IEEEnonumber
\\
&&\hspace{.2cm}\ell^{\ast}_{\text{sr}}(r,\phi):=\frac{1+|\xs-\xr|^{\alpha}}{1+(r^2+\xr^2-2r\xr\cos\phi)^{\frac{\alpha}{2}}}\IEEEnonumber
			\end{IEEEeqnarray}
     and assume $|\xs|\neq|\xr|$. For exponentially distributed $\mathsf{u}$, $\mathsf{g}$ and $\mathsf{h}$, the OPs $q_{\text{BC}}$ and $q_{\text{MAC}}$ are given by
  \begin{IEEEeqnarray}{rCl}
  q_{\text{BC}}&=&\hspace{-0.02cm}1\hspace{-0.02cm}-\hspace{-0.02cm}\exp\hspace{-0.06cm}\big\{\hspace{-0.08cm}-\hspace{-0.06cm}\lambda\Psi\hspace{-0.02cm}\big(0,\tfrac{\beta}{2}\ell^{\ast}_{\text{sd}}(r)\big)\hspace{-0.06cm}\big\}\hspace{-0.06cm}-\hspace{-0.02cm}\exp\hspace{-0.06cm}\big\{\hspace{-0.08cm}-\hspace{-0.06cm}\lambda\Psi\hspace{-0.02cm}\big(\beta\ell^{\ast}_{\text{sr}}(r,\phi),0\big)\hspace{-0.06cm}\big\}\hspace{-0.06cm}\IEEEnonumber\\
		&&+\exp\hspace{-0.06cm}\big\{\hspace{-0.08cm}-\lambda\Psi\big(\beta\ell^{\ast}_{\text{sr}}(r,\phi),\tfrac{\beta}{2}\ell^{\ast}_{\text{sd}}(r)\big)\hspace{-0.06cm}\big\},\label{eq:op_rayray_bc}\IEEEeqnarraynumspace
  \end{IEEEeqnarray}
     and 
  \begin{IEEEeqnarray}{rCl}
		q_{\text{MAC}}&=&\exp\big\{-\lambda\Psi\big(\beta\ell^{\ast}_{\text{sr}}(r,\phi),0\big)\big\}\IEEEnonumber
    \\&&-\mu_{1}\exp\big\{-\lambda\Psi\big( \beta\ell^{\ast}_{\text{sr}}(r,\phi),\beta\ell^{\ast}_{\text{sd}}(r)\big)\big\}\IEEEnonumber\\
		&&+\mu_{2}\exp\big\{-\lambda\Psi\big( \beta\ell^{\ast}_{\text{sr}}(r,\phi),\beta\ell^{\ast}_{\text{rd}}(r)\big)\big\},\IEEEeqnarraynumspace\label{eq:op_rayray_mac}
	\end{IEEEeqnarray}
	where
	%\begin{IEEEeqnarray}{c}
		$\mu_{1}=\frac{\ell(|\xs|)}{\ell(|\xs|)-\ell(|\xr|)}$, $\mu_{2}=\frac{\ell(|\xr|)}{\ell(|\xs|)-\ell(|\xr|)}$, and
  %\begin{IEEEeqnarray}{rCl}
   $\Psi(f,g)=\int_{0}^{\infty}\int_{0}^{\pi}2r\,\big(1-\frac{1}{(1+f(r,\phi))(1+g(r))}\,\big)\,\mathrm d\phi\,\mathrm dr$.%.\IEEEeqnarraynumspace\label{eq:op_rayray_psi}
	%\end{IEEEeqnarray}
\end{proposition}
\begin{IEEEproof}
	We follow the approach used in \cite{baccelli06}: we first condition \eqref{eq:op_adaf} on $\Phi$ and evaluate the probabilities w.r.t. $\mathsf{u}$. Note that for $\ell(|\xs|)\neq \ell(|\xr|)$, the sum $\mathsf{z}=\mathsf{u}_{\text{sd}}\ell(|\xs|)+\mathsf{u}_{\text{rd}}\ell(|\xr|)$ has distribution
	\begin{IEEEeqnarray}{rCl}
		\mathbb{P}\left(\mathsf{z}>z\right)&=&\frac{\ell(|\xr|)e^{-z\ell(|\xr|)}}{\ell(|\xs|)-\ell(|\xr|)}-\frac{\ell(|\xs|)e^{-z\ell(|\xs|)}}{\ell(|\xs|)-\ell(|\xr|)}. \IEEEeqnarraynumspace
\end{IEEEeqnarray}
We then de-condition on $\Phi$ and exploit the linearity property of the expectation. We apply the definition of the Laplace transform for Poisson shot-noise processes with independent marks \cite{baccelli09} and insert the intensity measure from \eqref{eq:intensity_measure}. %, i.e.,
%\begin{IEEEeqnarray}{rCl}
%		&&\mathbb{E}\left[e^{-s (a\mathsf{Y}_{\text{r}}+b\mathsf{Y}_{\text{d}})}\right]=\mathbb{E}\left[\prod_{i\in\Phi}e^{-sa\mathsf{g}_{i}\ell(|\mathsf{x}_{i}-\xr|)-sa\mathsf{h}_{i}\ell(|\mathsf{x}_{i}|)}\right]\IEEEeqnarraynumspace\\
%		&&\hspace{.2cm}=1-\exp\left[-\lambda \int_{\mathbb{R}^2}1-\left(\int_{\mathbb{R}_{+}}e^{-sag\ell(|x-\xr|)}\,\mathrm dg\right.\right.\IEEEnonumber\\
%		&&\hspace{3.7cm}\left.\left.\times\int_{\mathbb{R}_{+}}e^{-sah\ell(|x|)}\,\mathrm dh\right)\mathrm dx\right].\IEEEeqnarraynumspace
%\end{IEEEeqnarray}
Using the fact that $\mathsf{g}$ and $\mathsf{h}$ are exponentially distributed, and switching to polar coordinates yields the result.
\end{IEEEproof}

\begin{remark}
  The OP $q$ for the case $|\xs|=|\xr|$ can be computed straightforward using a similar approach, see e.g., \cite{altieri12}. Due to space limitations we do not present this result here. 
\end{remark}

\subsubsection{Diversity order analysis}
We begin our analysis by noting the following Lemma.\footnote{Some notation: $\hspace{-0.04cm}f(z)\hspace{-0.05cm}\widesim[1.5]{z\to0}\hspace{-0.05cm}g(z)$ means $\lim_{z\to0}f(z)/g(z)=c$, $0\hspace{-0.04cm}<\hspace{-0.04cm}c\hspace{-0.04cm}<\hspace{-0.04cm}\infty$. We use $b(x,r)$ to denote a two-dimensional ball of radius $r$ centered at $x$.}
\begin{lemma}\label{lem:exp_scaling}
	Let $w(t)=\sum_k  a_{k}\left(1-e^{-t z_{k}}\right)$, where $t\geq0$ and $a_{k},z_{k}\in\mathbb{R}$. Then, $w(t)\widesim[1.5]{t\to0}t$ if and only if $\sum_k a_k z_k \neq 0$.
\end{lemma}
\begin{IEEEproof}
	By the power series $e^{-t z}=\sum_{k=0}^{\infty}\frac{ (-t z)^{k}}{k!}$, we can rewrite $w(t)$ as
	\begin{IEEEeqnarray}{c}
		w(t)=t\sum_k a_k z_k-\frac{t^2}{2}\sum_k a_k z_k^2+\ldots\label{eq:proof_lemma1}
	\end{IEEEeqnarray}
	showing that the first order coefficient in \eqref{eq:proof_lemma1} must be non-zero to obtain the desired scaling.
\end{IEEEproof}
We are now in the position to derive the first results.
\begin{theorem}\label{thm:rayray_do}
	The achievable SC-DO of SDF for exponentially distributed $\mathsf{u}$, $\mathsf{g}$ and $\mathsf{h}$ is $\Delta=1$. 
\end{theorem}
A proof is given in Appendix~\ref{ap:proof_rayray_do}. Theorem~\ref{thm:rayray_do} states that there is no SC-DO gain by relaying the source's packet---which is a negative result since it is known that SDF achieves diversity order of two in the interference-free case \cite{laneman04}. This pitfall results from the fact that by simply forwarding the source's packet, the relay cannot change the interference level at the destination in the second time slot ($q_{\text{MAC}}\widesim[1.5]{\lambda\to0}\lambda$). 

On the other hand, spatial correlation of the interference, compounded with channel fading, renders the BC phase not as effective as in the interference-free case ($q_{\text{BC}}\widesim[1.5]{\lambda\to 0}\lambda$). Increasing the relay-destination separation can lower this undesirable effect and provide a stronger scaling of $q_{\text{BC}}$ at intermediate $\lambda$, but it cannot steepen the asymptotic slope of $q_{\text{BC}}$.  

For these reasons, there is no reliability gain w.r.t.\ the interference. Yet the relay can provide a power gain compared to direct transmission.% by increasing the received signal power at the destination.

\subsubsection{Optimal relay position}
Theorem~\ref{thm:rayray_do} and the ensuing discussion give rise to the question about the optimal relay position. Using Proposition~\ref{prop:rayray_op}, we are able to numerically minimize $q$ over $\xr$ given $\xs$.

\begin{figure}[t]
  \psfrag{tag1}[c][c]{\small{$\alpha$}}
  \psfrag{tag2}[c][c]{\small{Optimal $\frac{|\xs-\xr|}{|\xs-\xd|}$}}
  \begin{center}
    \includegraphics[width=.47\textwidth]{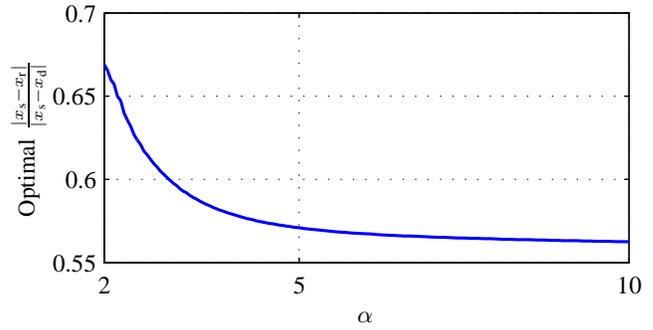}
  \end{center}
  \caption{Optimal relay position relative to source-destination distance as a function of the path loss exponent. Line configuration assumed.}
  \label{fig:d_opt}\vspace{-.3cm}
\end{figure}

Glancing at Fig.~\ref{fig:d_opt}, we make a surprising observation: it is better to put the relay closer to the destination (receive diversity) rather than to the source (transmit diversity), thereby showing an adverse behavior compared to the interference-free case. The intuition behind this observation is that the ability to boost the received power through the relay-destination link outweighs the reliability loss of the source-relay link. Motivated by this result, we next focus on the achievable SC-DO in the BC-phase only.  
\section{Diversity Analysis in BC-Phase}
In Section~\ref{sec:out_ray} it was concluded that the SC-DO is negatively affected mainly due to the invariability of the interference at the destination in the MAC-phase---the relay cannot provide diversity w.r.t. the interference. This invariability moreover does not change when a different fading distribution is assumed. In contrast, the SC-DO of only the BC-phase can theoretically be higher because the interference de-correlates over space. In the proof of Theorem~\ref{thm:rayray_do}, however, it was shown that for Rayleigh fading $q_\text{BC}\widesim[1.5]{\lambda\to0}\lambda$ unless $|\xd-\xr|\to\infty$. Since this result relies on the interplay between the spatial interference correlation and exponentially distributed fading gains, we next study the achievable SC-DO in the BC-phase for different assumptions about the fading.

\subsection{Non-fading links + fading interference}\label{sec:non_fad_and_fad}

In many scenarios CSI is available at the transmitter, typically indicating the instantaneous channel gains of the desired links ($\mathsf{u}$) to which $\beta$ can then be adapted. The instantaneous channel gains of the interfering links ($\mathsf{g},\mathsf{h}$) however usually remain unknown to the transmitter due to practical constraints. In what follows, we modify our model by conditioning \eqref{eq:op_adaf} on $\mathsf{u}$, thereby noting that outages are now due to interference only. The \emph{dominant interferer} phenomenon (cf. \cite{weber10}) will play an important role for the derivation of the subsequent results.

\begin{definition}\label{def:dominant_def}
  An interferer is called dominant if its individual interference contribution is already sufficiently high to create outage. The set of dominant interferers at the relay (destination) is defined as $\tilde{\Phi}_{\text{r}}\subseteq\Phi$ ($\tilde{\Phi}_{\text{d}}\subseteq\Phi$).
\end{definition}

\begin{proposition}\label{prop:op_no_fad_fad}
  The OP $q_{\text{BC}}$ for SDF in the case of non-fading links ($\mathsf{u}\equiv1$) and fading interference is lower bounded by 
  \begin{IEEEeqnarray}{rCl}
    &&q_{\text{BC}}\geq1-\exp\Big\{-2\lambda\int_{\mathbb{R}_{+}}r\,\mathbb{P}\big(\mathsf{h}>\ell_{\text{sd}}^{\ast}(r)^{-1}\tfrac{2}{\beta}\big)\IEEEnonumber\\
    &&\hspace{2.5cm}\times\int_{0}^{\pi}\mathbb{P}\big(\mathsf{g}>\ell_{\text{sr}}^{\ast}(r,\phi)^{-1}\tfrac{1}{\beta}\big)\,\mathrm d\phi\,\mathrm dr\Big\}.\IEEEeqnarraynumspace\label{eq:non_fad_fad}
  \end{IEEEeqnarray}

\end{proposition}
A proof is given in Appendix~\ref{proof:op_no_fad_fad}. We are now able to analyze the achievable SC-DO for this case.
\begin{theorem}\label{thm:fad_no_fad}
  The achievable SC-DO of SDF in the BC-phase for the case of non-fading links and fading interference is $\Delta=1$.
\end{theorem}
\begin{IEEEproof}
  Note that the expressions under the integral signs in \eqref{eq:non_fad_fad} are always positive. By Lemma~\ref{lem:exp_scaling} and since \eqref{eq:non_fad_fad} is a lower bound, we thus have $q_{\text{BC}}\widesim[1.5]{\lambda\to0}\lambda$.
\end{IEEEproof}

\subsection{Path loss only model}\label{sec:non_fad_links}
In case of weak scatterings between all nodes, the channel can be characterized by the path loss only model, for which an asymptotically tight lower bound on the interference tail probability exists for the path loss law $r^{-\alpha}$. The next Lemma extends this statement to the non-singular path loss law.\footnote{To the best of the authors' knowledge, the statement in Lemma~\ref{lem:tightness} was not found explicitly in the literature.}
\begin{lemma}\label{lem:tightness}
  The interference tail probability lower bound based on the dominant interferer phenomenon is tight as $\lambda\to0$ also for the non-singular path loss law defined in Section~\ref{sec:system_model}.
\end{lemma}

\begin{IEEEproof}
  We first note that the interference tail probability lower bound based on the maximum interferer principle for the path loss law $r^{-\alpha}$ is tight \cite{weber10}. This is due to the singularity at $r=0$ which renders the interference sub-exponential---by allowing the maximum of the individual interference contributions to dominate the sum interference. In contrast, the interference in our case is not sub-exponential because of the boundedness of our path loss function. However, it is intuitive that the interference is nevertheless able to ``make a single relatively big jump'' whenever the close neighborhood of the receiver carries no (statistical) weight; which is the case at small $\lambda$. Indeed, denoting by $r_{n}$ the $n$-th nearest interferer, 
\begin{IEEEeqnarray}{rCl}
	\mathbb{P}\left(r_{n}^{-\alpha}(1+r_{n}^{\alpha})>1+\epsilon\right)&=&%\mathbb{P}\left(r_{n}<\epsilon^{-\frac{1}{\alpha}}\right)\IEEEeqnarraynumspace\IEEEnonumber\\
  1-\frac{\Gamma(n,\lambda\pi\epsilon^{-\frac{2}{\alpha}})}{\Gamma(n)}\overset{\lambda\to0}{\longrightarrow}0,\IEEEeqnarraynumspace\label{eq:proof_nearest}
\end{IEEEeqnarray}
where $\Gamma(a,z)=\int_{z}^{\infty}t^{a-1}e^{-t}\,\mathrm dt$ is the upper incomplete Gamma function. Thus, \eqref{eq:proof_nearest} states that the individual contributions of the $n$-nearest interferers become equal for the two path loss laws as $\lambda\to0$. From this equivalence it follows that, in the small density regime, the dominance of the nearest interferer is preserved with our path loss law. This, in turn, renders the dominant-interferer based lower bound tight: when the nearest interferer is not a member of the dominant set ($\tilde{\Phi}=\emptyset$) it is likely that no outage occurs since adding the sum interference from the remaining interferers to the maximum interference will most likely not deteriorate the SIR much.
\end{IEEEproof}

Using the fact that the dominant-interferer bound is asymptotically tight, we are now able to study the SC-DO.

\begin{theorem}\label{thm:non_fad}
Define $\mathcal{A}_{\text{r}}:=b(\xr,r_{1})$, $\mathcal{A}_{\text{d}}:=b(\xd,r_{2})$ and $\mathcal{A}_{\text{r,d}}:=\mathcal{A}_{\text{r}}\cap\mathcal{A}_{\text{d}}$, where $r_{1}=(\beta\ell(|\xs-\xr|)^{-1}-1)^{1/\alpha}$ and $(\tfrac{1}{2}\beta\ell(|\xs|)^{-1}-1)^{1/\alpha}$.
% \begin{IEEEeqnarray}{C}
% 	\mathcal{A}_{\text{r}}:=b\left(\xr,\Big(\frac{\beta}{\ell(|\xs-\xr|)}-1\Big)^{2/\alpha}\right)\label{eq:dom_reg1}\\
% 	\mathcal{A}_{\text{d}}:=b\left(\xd,\Big(\frac{\beta}{2\ell(|\xs|)}-1\Big)^{2/\alpha}\right),\label{eq:dom_reg2}
% \end{IEEEeqnarray}
% which are the dominant interference regions around the relay and the destination, respectively, and let $\mathcal{A}_{\text{r,d}}:=\mathcal{A}_{\text{r}}\cap\mathcal{A}_{\text{d}}$.
Then, $q_{\text{BC}}$ for SDF in the path loss only case ($\mathsf{u}\equiv\mathsf{g}\equiv\mathsf{h}\equiv1$) is given by

\begin{numcases}{q_{\text{BC}}\widesim[1.5]{\lambda\to0}}
\lambda|\mathcal{A}_{\text{r,d}}|,	&	$\mathcal{A}_{\text{r,d}}\neq\emptyset$\\
\lambda^2|\mathcal{A}_{\text{r}}|\,|\mathcal{A}_{\text{d}}|, & $\mathcal{A}_{\text{r,d}}=\emptyset$,
\end{numcases}
and the achievable SC-DO in the BC-phase is
\begin{numcases}{\Delta=}
1,& $|\xr-\xd|\leq r_{1}+r_{2}$\IEEEeqnarraynumspace\\
2,& otherwise.
\end{numcases}
\end{theorem}

\begin{figure}[t]
  \psfrag{tag1}[c][c]{\small{$\lambda$}}
  \psfrag{tag2}[c][c]{\small{$q_{\text{BC}}$}}
  \psfrag{tag3}{\small{No relay}}
  \psfrag{tag4}{\small{$|\xr|/|\xs|=.40$}}
  \psfrag{tag5}{\small{$|\xr|/|\xs|=.53$}}
  \psfrag{tag6tag6tag6t}{\small{$|\xr|/|\xs|=.66$}}
  \psfrag{tag7}[c][c]{$|\mathcal{A}_{\text{r,d}}|\approx 23.0\%$}
  \psfrag{tag8}[c][c]{$|\mathcal{A}_{\text{r,d}}|\approx 7.7\%$}
  \psfrag{tag9}[c][c]{$|\mathcal{A}_{\text{r,d}}|\approx 0\%$}
    \includegraphics[width=0.48\textwidth]{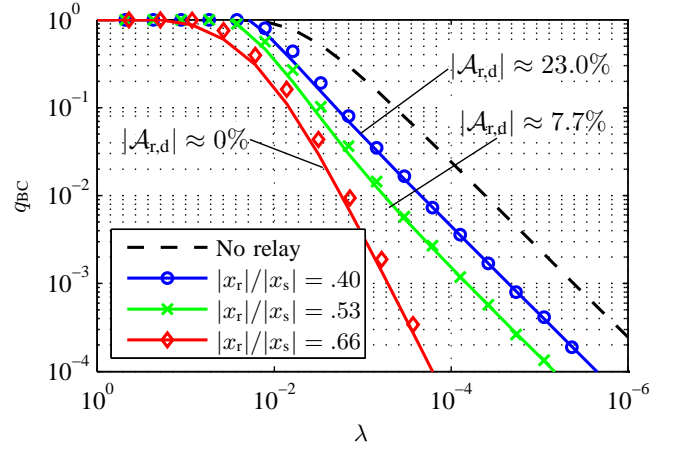}
    \caption{OP $q_{\text{BC}}$ vs. $\lambda$. System parameters are: $\xs=(15,0)$, $\alpha=4$, $\beta=.1$. Marks represent simulation results.}
  \label{fig:pout_ppl}\vspace{-.3cm}
\end{figure}

\begin{IEEEproof}
The proof is similar to the one of Proposition~\ref{prop:op_no_fad_fad}: we first re-define the dominant interferer sets for our purposes, leading to the regions $\mathcal{A}_{\text{r}}$, $\mathcal{A}_{\text{r}}$ and their intersection $\mathcal{A}_{\text{r,d}}$. By Lemma~\ref{lem:tightness}, we then have
\begin{IEEEeqnarray}{rCl}
q_{\text{BC}}&\widesim[1.5]{\lambda\to0}&\mathbb{P}\big(\Phi(\mathcal{A}_{\text{r,d}})\neq\emptyset\big)\IEEEnonumber\\
&&+\mathbb{P}\big(\Phi(\mathcal{A}_{\text{r}}\setminus\mathcal{A}_{\text{d}})\neq\emptyset\big)\,\mathbb{P}\big(\Phi(\mathcal{A}_{\text{d}}\setminus\mathcal{A}_{\text{r}})\neq\emptyset\big),\IEEEeqnarraynumspace
\end{IEEEeqnarray}
where we make use of the independence property of the PPP\@. Using the fact $1-\exp(-\lambda z)\widesim[1.5]{\lambda\to0}\lambda z$ yields the result.
\end{IEEEproof}
The regions $\mathcal{A}_{\text{r}}$ and $\mathcal{A}_{\text{r}}$ as well as their intersection play a crucial role for the resulting diversity behavior: whenever there is no overlapping of the individual dominant-interferer regions, the interference at the relay and at the destination can be assumed independent as $\lambda\to0$, yielding $\Delta=2$. The fact that the transition from $\Delta=1$ to $\Delta=2$ is not continuous might seem counter-intuitive first; as long as there is a non-zero probability for the occurrence of a jointly-dominant interferer ($\mathcal{A}_{\text{r,d}}\neq\emptyset$), the linear term will be dominant as $\lambda\to0$. Simulations confirm this result as can be seen in Fig.~\ref{fig:pout_ppl}.

\begin{remark}
	Theorem~\ref{thm:rayray_do} and Theorem~\ref{thm:fad_no_fad} also hold for the case where the interferers perform SDF as well. Assuming synchronous transmissions, this can be checked by regarding the interference power of the interfering relays as being created by the corresponding interfering source nodes and undergoing a \emph{modified} fading distribution.
\end{remark}

\section{Conclusion}
Using point process theory and a modification of the diversity order metric suitable for interference-limited networks, our analysis reveals that the achievable \emph{spatial-contention} diversity order (SC-DO) of selection decode-and-forward is equal to one. This is because conventional decode-and-forward relaying, in general, cannot reduce the interference at the destination. As a consequence, the relay should be placed closer to the destination (receive diversity) to provide considerable power boosts. The analysis shows that such a receive-diversity configuration is better in terms of achievable SC-DO: depending on the interference correlation between relay and destination, an SC-DO of two is achievable when fading is negligible and the relay-destination link is reliable. The insights obtained may be of interest for designing cooperative receive-diversity techniques for contemporary wireless networks. A possible extension could be to further study the achievable diversity order for the case of non-Poisson interference, e.g., when the interferers perform cooperative relaying as well.

\vspace{-0.00cm}
\section*{Acknowledgements}
The authors gratefully acknowledge that their work is partially
supported within the priority program 1397 "COIN" under grant No. JO
258/21-1 by the German Research Foundation (DFG).


\appendices
\vspace{-0.22cm}
\section{Proof of Theorem~\ref{thm:rayray_do}}\label{ap:proof_rayray_do}
Taking the limit $\lambda\to 0$ in \eqref{eq:op_rayray_bc} and \eqref{eq:op_rayray_mac}, we obtain
\begin{IEEEeqnarray}{rCl}
	q&\widesim[1.5]{\lambda\to0}&\lambda\,\Big[-\Psi\big(\beta\ell^{\ast}_{\text{sr}}(r,\phi),\tfrac{\beta}{2}\ell^{\ast}_{\text{sd}}(r)\big)\IEEEnonumber\\
	&&\hspace{1.2cm}+\Psi\big(0,\tfrac{\beta}{2}\ell^{\ast}_{\text{sd}}(r)\big)+\Psi\big(\beta\ell^{\ast}_{\text{sr}}(r,\phi),0\big)\Big]\IEEEnonumber\\
	&&\hspace{.0cm}+\lambda\,\Big[-\Psi\big(\beta\ell^{\ast}_{\text{sr}}(r,\phi),0\big)+\mu_{1}\Psi\big( \beta\ell^{\ast}_{\text{sr}}(r,\phi),\beta\ell^{\ast}_{\text{sd}}(r)\big)\IEEEnonumber\\
	&&\hspace{1.2cm}-\mu_{2}\Psi\big( \beta\ell^{\ast}_{\text{sr}}(r,\phi),\beta\ell^{\ast}_{\text{rd}}(r)\big)\Big]+\mathcal{R}(\lambda)
	,\IEEEeqnarraynumspace\label{eq:proof_rayray_op1}
\end{IEEEeqnarray}
where $\mathcal{R}(\lambda)$ contains all non-linear terms. By Lemma~\ref{lem:exp_scaling}, the linear term of $q$ must be non-vanishing for the Theorem to hold. Thus, we only need to prove that the linear term is non-zero, for which strictly positiveness of the expressions inside the two brackets is a sufficient condition. In what follows, we will prove that the strictly-positiveness condition is fulfilled for each of them. For each of the two expressions, we insert $\Psi(\cdot,\cdot)$ and rewrite the sum of integrals by a single one, thereby exploiting the linearity property of integrals. A sufficient condition for strictly-positiveness of the two integrals is when their integrands are strictly positive almost everywhere. After some algebraic manipulations, we therefore have to check if
\begin{IEEEeqnarray}{rCl}
	&&1+\frac{1}{(1+\tfrac{\beta}{2}\ell^{\ast}_{\text{sd}}(r))(1+\beta\ell^{\ast}_{\text{sr}}(r,\phi))}\IEEEnonumber\\
	&&\hspace{1.3cm}-\frac{1}{1+\tfrac{\beta}{2}\ell^{\ast}_{\text{sd}}(r)}-\frac{1}{1+\beta\ell^{\ast}_{\text{sr}}(r,\phi)}>0\IEEEeqnarraynumspace\label{eq:proof_rayray_op2}
\end{IEEEeqnarray}
for the $q_{\text{BC}}$-part and
\begin{IEEEeqnarray}{rCl}
	&&\mu_{2}(1+\beta\ell^{\ast}_{\text{sd}}(r))-\mu_{1}(1+\beta\ell^{\ast}_{\text{rd}}(r))\IEEEnonumber\\
	&&\hspace{1.7cm}+(1+\beta\ell^{\ast}_{\text{rd}}(r))(1+\beta\ell^{\ast}_{\text{sd}}(r))>0\IEEEeqnarraynumspace\label{eq:proof_rayray_op3}
\end{IEEEeqnarray}
for the $q_{\text{MAC}}$-part. Both \eqref{eq:proof_rayray_op2} and \eqref{eq:proof_rayray_op3} are readily shown to be strictly positive, implying that the linear term of $q$ is strictly positive as well. This proves the result.\qed

\section{Proof of Proposition~\ref{prop:op_no_fad_fad}}\label{proof:op_no_fad_fad}
We start by formalizing the definition of dominant sets:
\begin{IEEEeqnarray}{c}
  \tilde{\Phi}_{\text{r}}:=\left\{\mathsf{x}_{i}\in\Phi:\frac{\mathsf{g}_{i}\ell(|\mathsf{x}_{i}-\xr|)}{\ell(|\xs-\xr|)}>\frac{1}{\beta}\right\}\IEEEeqnarraynumspace\label{eq:proof_op_no_fad_fad_1}\\
  \tilde{\Phi}_{\text{d}}:=\left\{\mathsf{x}_{i}\in\Phi:\frac{\mathsf{h}_{i}\ell(|\mathsf{x}_{i}|)}{\ell(|\xs|)}>\frac{2}{\beta}\right\}\IEEEeqnarraynumspace\label{eq:proof_op_no_fad_fad_2}
\end{IEEEeqnarray}
and
\begin{IEEEeqnarray}{c}
  \tilde{\Phi}_{\text{r,d}}:=\left\{\mathsf{x}_{i}\in\Phi\hspace{-.03cm}:\frac{\mathsf{g}_{i}\ell(|\mathsf{x}_{i}-\xr|)}{\ell(|\xs-\xr|)}\hspace{-.03cm}>\hspace{-.03cm}\frac{1}{\beta}\,\land\,\frac{\mathsf{h}_{i}\ell(|\mathsf{x}_{i}|)}{\ell(|\xs|)}\hspace{-.03cm}>\hspace{-.03cm}\frac{2}{\beta}\right\}\hspace{-.04cm}.
  \IEEEeqnarraynumspace\label{eq:proof_op_no_fad_fad_3}
\end{IEEEeqnarray}
Note that since $\tilde{\Phi}_{\text{r,d}}=\tilde{\Phi}_{\text{r}}\cap\tilde{\Phi}_{\text{d}}$, the occurrence of the event $\{\tilde{\Phi}_{\text{r,d}}\neq\emptyset\}$ is a sufficient condition for $\{\tilde{\Phi}_{\text{r}}\neq\emptyset\,\land\,\tilde{\Phi}_{\text{d}}\neq\emptyset\}$. Therefore, we have $\mathbb{P}( \tilde{\Phi}_{\text{r,d}}\neq\emptyset)\leq\mathbb{P}(\tilde{\Phi}_{\text{r}}\neq\emptyset\,\land\,\tilde{\Phi}_{\text{d}}\neq\emptyset)$. Thus,
\begin{IEEEeqnarray}{rCl}
  q_{\text{BC}}&=&\mathbb{P}\left(\sum_{i\in\Phi}\frac{\mathsf{g}_{i}\ell(|\mathsf{x}_{i}-\xr|)}{\ell(|\xs-\xr|)}>\frac{1}{\beta},\sum_{i\in\Phi}\frac{\mathsf{h}_{i}\ell(|\mathsf{x}_{i}|)}{\ell(|\xs|)}>\frac{2}{\beta}\right)\IEEEnonumber\\
  &\geq&\mathbb{P}\left(\sum_{i\in\tilde{\Phi}_{\text{r}}}\frac{\mathsf{g}_{i}\ell(|\mathsf{x}_{i}-\xr|)}{\ell(|\xs-\xr|)}>\frac{1}{\beta},\sum_{i\in\tilde{\Phi}_{\text{d}}}\frac{\mathsf{h}_{i}\ell(|\mathsf{x}_{i}|)}{\ell(|\xs|)}>\frac{2}{\beta}\right)\IEEEnonumber\\
  &=&\mathbb{P}\left(\tilde{\Phi}_{\text{r}}\neq\emptyset,\tilde{\Phi}_{\text{d}}\neq\emptyset\right)\IEEEnonumber\\
  &\geq&\mathbb{P}\left(\tilde{\Phi}_{\text{r,d}}\neq\emptyset\right)=1-\exp\left(-\psi\right),
  \IEEEeqnarraynumspace\label{eq:proof_op_no_fad_fad_4}
\end{IEEEeqnarray}
where the last equality follows from the number of elements in $\tilde{\Phi}_{\text{r,d}}$ being Poisson distributed with mean $\psi$, which can be computed straightforward using \eqref{eq:intensity_measure} as
\begin{IEEEeqnarray}{rCl}
  &&\psi=\lambda\int_{\mathbb{R}^2}\,\mathbb{P}\left(\mathsf{g}>\frac{\ell(|\xs-\xr|)}{\beta\ell(|x-\xr|)}\right)\mathbb{P}\left(\mathsf{h}>\frac{2\ell(|\xs|)}{\beta\ell(|x|)}\right)\,\mathrm dx.
  \IEEEeqnarraynumspace\label{eq:proof_op_no_fad_fad_5}
\end{IEEEeqnarray}
This concludes the proof.\qed

\bibliographystyle{IEEEtran}
%\bibliography{IEEEabrv,../../literature}

\end{document}